\documentclass[twocolumn]{emulateapj}
\submitted{Science with Parkes @ 50 Years Young, 31 Oct. -- 4 Nov., 2011}

\shorttitle{Early Parkes Observations}
\shortauthors{K. I. Kellermann}

\begin{document}


\title{Early Parkes Observations of Planets \\
    and Cosmic Radio Sources}


\author{K. I. Kellermann}
\affil{National Radio Astronomy Observatory, Charlottesville, VA 22901}
\email{kkellerm@nrao.edu}



\begin{abstract}
We discuss early Parkes observations of the radio emission from the planets Mercury, Venus, Mars, Saturn, and Uranus.  The sensitive Parkes 11 cm system was used to detect a surprisingly high observed nighttime temperature on Mercury, the first, but unrecognized, hint that the Mercury actually rotates with respect to the Sun, as well as detecting the faint radio emission from Uranus. We also discuss the anomalous spectrum of PKS 1934-63, the first recognized GPS source.
 
\end{abstract}





\section{Introduction}

When the Parkes radio telescope went into operation, it was so much more powerful than any thing else that existed at the time, that one did not need to be an expert about anything to just point the telescope and make new discoveries.

I came to the Radiophysics Lab (RP) in 1963 with the intention of using the Parkes radio telescope to continue the work I did for my PhD thesis on radio source spectra.  After a few months, I moved to Parkes, since that is where all the action was.  John Bolton was the Director and ran the observatory with an iron hand.  He made all the rules, and there was no room for argument or negotiation.  One of his rules was that the telescope could be used for observing only at night, and the days were reserved for maintenance and testing.  Often, I noticed that there was no maintenance or testing going on, but the telescope was sitting idle.  It really bothered me to see this fantastic instrument just sitting there unused.  But as much as I pleaded, John would not compromise his principles.

It wasn't just a matter of finding a telescope operator, or driver, as they were called, since I was a licensed driver myself, having been suitably trained and tested by John himself.  The only exception to his ``nighttime only'' policy, he explained, was to observe something that could not be observed in the night sky.  After careful thought, I concluded that this meant the occasional supernova, the Sun, and Mercury.   I wasn't prepared to wait around a few hundred years for the next supernova, and I wasn't interested in the Sun.  Beside the Sun was too complex with all those different types of bursts, and anyway there was a whole solar division at RP that studied the Sun.  That left just Mercury.

\section{The Planets}

I didn't know very much about the planets, but I learned that they are heated by the Sun, and that their surface temperature depends mostly on their distance from the Sun, but also their albedo (how much energy is reflected rather than absorbed), and the rate that the planet rotates (absorbs and radiates heat). 

\subsection{Mercury}

All the astronomy text books of the time said that since the rotation period of Mercury is the same as its period of revolution around the Sun, that there was perpetual day time on one hemisphere, and perpetual night on the other.  This generally accepted idea apparently dated from the 1888 work of Giovanni Schiaparelli \citep{H90,L02} on the basis of a few observations and his line drawings of surface features that he imagined seeing.

Typically, Celia-Payne Gaposhkin's 1954  text book, \it Introduction to Astronomy, \rm reported \it The sun beats fiercely on the face of Mercury; 
 ... Mercury always turns the same face to the sun, ... 
and so one side of the planet is in continual sunlight;  the other is in perpetual  shadow.   Under the sun's  rays the surface of Mercury is kept at a temperature  near 350 C ....  But the other side of the planet is in eternal darkness, and its temperature cannot be far from absolute zero \citep{PG54}.

\rm

So, I knew if I wanted to study Mercury at Parkes, I would need to observe when the planet was at superior conjuction on the other side of the Sun when the hot day lit side would be facing the Earth.  I expected that the signal would be weak, and that it would be necessary to average many scans to detect any radio emission from Mercury.  To complicate the situation, Mercury is always near the Sun, so that the side lobes of the strong solar emission can easily be stronger than the weak thermal emission from Mercury.  

There had been only one previous measurement of the radio emission from Mercury, by \citet{WEH62} which was made when approximately half of the illuminated planet faced the Earth.  \citet{WEH62} measured a surprisingly high average disk temperature of about 400 K.  Assuming the nighttime  temperature was close to absolute zero, they calculated a subsolar (noon time) temperature of 1100$\pm$300 K, considerably higher than expected from solar heating.

Since the  signal from Mercury was expected to be weak, I rigged up a system that John Bolton had developed to digitize the telescope output which was sampled every few seconds, and then written on paper by an electric typewriter. To reduce the noise, I planned to average the numbers at each point from multiple scans using an electric adding machine. In order to gain experience in dealing with the solar side lobes, I started to observe just as Mercury was passing inferior conjunction when only the presumably cold dark side of the planet would be available. Much to my surprise Mercury came up loud and clear on a single scan.  The apparent disk temperature was about 300 K, close to room temperature.  I followed the planet as it revolved around the sun so that the sunlit side became more and more exposed.  I wasn't sure what to expect.  Would the sunlit side also be hotter than the night side?  No, the apparent disk temperature remained nearly constant, meaning that the day and night sides must have about the same surface temperature.

How could this be?  The text books were clear on two things.  Mercury doesn't rotate and doesn't have an atmosphere to circulate heat from the sunlit to the dark side.  The evidence for non rotation appeared to me to be stronger than the evidence for no atmosphere.  In fact, George \cite{F62, F64} had earlier suggested that Mercury might, in fact, contain a thin Argon atmosphere and predicted a dark side temperature between 50 K and 250 K.  So I interpreted my anomalous temperature observations in terms of an atmosphere that could sustain the propagation of warm air from the daytime to the nighttime side.

The true interpretation was understood just a few months later, when radar measurements showed unambiguously that Mercury in fact did rotate with a 59 day period, precisely 2/3 of the rate of revolution around the sun \citep{PD65}.  Characteristically,  the theoreticians were quick to point out that this value was close to that expected on theoretical grounds \citep{C65,PG65} although until the radar observations, they were apparently comfortable with an  88 day synchronous period. Apparently, Schaparelli, and others after him, had been fooled into believing that the rotation and revolution periods of Mercury were identical, because of the 3:2 spin-orbit resonance so that the synodic period is twice the rotation period; thus the same side of Mercury faces the Earth whenever Mercury is best observed in its very eccentric orbit at maximum distance from the Sun.  Had I understood this, and had not been convinced by all the text books that Mercury doesn't rotate, I might have been the first to understand that the warm nighttime temperture measured at 11 cm is due direct solar heating resulting from the rotation of the planet.  

I learned from this experience that one should always go back to the original work, and not to trust text books or other secondary literature.  Unfortunately I don't read French, so I didn't realize that Schaparelli's data was so flaky, and apparently neither had any of the many text book writers for nearly the next hundred years.

\subsection{Mars}

At the 1964 IAU General Assembly in Hamburg, Rod Davies announced that he had measured a surprisingly high Martian temperature of 1,140$\pm$50K at 21 cm using the Jodrell Bank 250-ft MK I radio telescope \citep{D64}, although earlier observations had reported an apparent disk temperature of only 211$\pm$20K at 3.15 cm and 177$\pm$17K at 10 cm made at the Naval Research Laboratory \citep{MMS58} and at NRAO \citep{H63} respectively.  These short wavelength values were close to that expected from solar heating, so Davies interpreted his observations in terms of intense non thermal radiation from charged particles circling the planet much like the Van Allen radiation belts observed a few years earlier around Jupiter by \citet{RR60}.  

This was a serious matter, since NASA was about to send a spacecraft to Mars.  If there really were belts of charged particles around Mars, they could affect the delicate instrumentation abord the spacecraft.  I was suspicious of the Jodrell Bank results since, the reported Martian flux density of only 0.2 Jy was comparable with the expected confusion noise of the MK I telescope at 21 cm.  So I used the Parkes antenna to measure the disk temperature of Mars at 6, 11, and 21 cm.  I pointed the antenna at the position of Mars on a number of successive days and measured the corresponding antenna temperature.  Then after Mars had moved by more than a beamwidth, I re measured the antenna temperature at each of the same positions.  The difference between the two measurements were then used to calculate the true antenna temperature due to Mars.  Indeed, Mars turned out to have a perfectly reasonable disk temperature  of 192 $\pm$ 26, 162 $\pm$ 18, and 169 $\pm$ 33 K at 6, 11, and 21 cm respectively \citep{K65}.  The Jodrell Bank measurements, were, indeed, thoroughly confused at 21 cm.

\subsection{The Other Planets}

By this time, I considered myself an expert on planetary radio astronomy, and set out to use the Parkes antenna to measure the surface temperature of the other planets. 

I observed Venus at 11, 21, 31, and 48 cm and was able to confirm the high temperature of Venus which had been previously measured at NRL \citep{MMS58}.  This excess temperature of Venus is now understood in terms of a Green House effect due to the heavy cloud cover which envelops the entire planet.  Although at the longer wavelengths, the effects of noise and confusion led to relatively large uncertainties, these new Parkes observations supported the somewhat lower disk temperature previously reported from observations at Arecibo \citep{H65} and Green Bank \citep{Dr64}.

Next I observed Saturn also at 6, 11, and 21 cm and measured disk temperatures of 179 $\pm$ 19, 196 $\pm$ 20, and 303 $\pm$ 50 K at 6, 11, and 21 cm respectively.  The increase in apparent temperature at the longer wavelengths was also speculated to be due to a Jupiter-like non thermal radiation belt.  However, subsequent high resolution observations showed no evidence for any radio emission beyond the planetary disk \citep{BR68}.  More likely the high temperature measured at the longer wavelengths is due to the higher temperature lower down in the planet's atmosphere which is probed by the longer wavelength radio emission.

Finally, I was able to use what was at the time, the most sensitive 11 cm system available anywhere to measure, for the first time, the thermal radiation from Uranus.  As the observed flux density of 8 mJy is well below the 11 cm confusion level of the Parkes telescope, the observations were made at multiple positions of the planet, and corrected for confusion by subtracting the observed flux density measured at each position after Uranus had moved outside of the antenna beam

\section{Radio source Spectra}

 When John Bolton came to Caltech, he and Gordon Stanley designed
and built the two element interferometer to measure accurate radio
positions in order to enable the optical identification of radio
sources.  In order to prepare a finding list free of 3C lobe shifts,
Dan Harris and Jim Roberts \citep{HR60} observed all of the 3C sources
using a single 90 foot element of the interferometer at a frequency of
960 MHz.  They observed using drift curves, and apparently on several
occasions Harris fell asleep at the controls and let the drift curve
continue for long after the source passed through the beam.  On two
such occasions, Dan acidently discovered a new source previously
unrecognized in the lower frequency catalogs.  He named theses CTA 21
and CTA 102.  As later shown by \citet{KLAM62} these were the first of
the category of what we now call Gigahertz Peaked Spectrum (GPS)
Sources.
 
 One of the first projects John Bolton initiated with the
new Parkes radio telescope was a dual frequency survey of the
available sky at 21 and 75 cm. John was aware of the unusual radio
spectra of CTA 21 and CTA 102 and was on the lookout for similar type
spectra.  The Parkes Survey was divided into four declination zones
and published in four installments.  Paper I \citep{BGM64} covered
$-20$ to $-60$ degrees declination; Paper II \citep{PM65} $-60$ to $-90$;
Paper III \citep{DSEC66} 0 to $+20$; and Paper IV \citep{SDEC66} 0 to
$-20$.  However, a careful inspection of the -$20$ to $-60$ catalog
\citep{BGM64} shows a surprising entry for the source 1934-63, which
is located at $-63$ degrees declination and therefore properly belonging
in the Price and Milne $-60$ to $-90$ degree catalog.
 
 As told by John
Bolton and Marc Price (private communications) what apparently
happened was that Marc had just finished an all night 14 hour run and
had left his chart records lying on the counter.  John, knowing about
CTA 21 and CTA 102 looked through Marc's records and immediately
noticed that 1934-643 was considerably stronger at 20 cm than at 75
cm.  He asked Marc,''Are you finished with these?''  Marc, being half
asleep from his all night vigil, replied that he indeed was finished.
The next Marc heard about 1934-63 was when it appeared in Nature
\citep{BGM63}.  In that same issue, on the facing page was the famous
paper by \citet{S63} which discussed the spectra of GPS sources in
terms of synchrotron self absorption (SSA).  In this paper, Slysh
showed that the presence of a low frequency spectral SSA cutoff
implies a very small angular size, and this was a large part of the
motivation a few years later, to develop VLBI.  Although the VLBI
observations were consistent with the predicted angular dimensions,
more recent studies suggest, in fact that the observed low frequency
cutoffs, may, in fact, be at least partially due to free-free (FFA)
and not SSA \citep{ LKV03}
 
 In order to try to distinguish between
SSA and FFA models, I used the Parkes telescope in early 1965 to
observe the spectrum of 1934-63 between 350 and 5000 MHz.  In the
critical region below 1 GHz, I used the standard Parkes simple crystal
mixer receiver and simple dipole feeds to observe at some 25
frequencies between 350 and 1000 MHz with spacings 20 to 25 MHz apart.
This system was reasonable broad band, and sensitivity was not an
issue with this strong source.  The effective observing frequency
could easily be changed by merely changing the local oscillator
frequency.  The problem was that the l.o. was located in the focal
cabin, and it took about half an hour to move the antenna to the
zenith, climb up to the focal cabin, then climb down, and return the
antenna to the 1934-63.  I only had one night of telescope time, and
clearly this would not work.  So I recruited Marc to spend the night
in the focal cabin to change the l.o. setting whenever I requested.  I
remember that it was a cold night, and shortly after midnight Marc
asked to be allowed to return to the control room to obtain a hair
dryer which he used to try to keep warm for the rest of the
night. Unlike Bolton, Garner, and Mackey, I gave Marc an
acknowledgment in my paper \citet{K66}.
 
 Years later, I asked John
Bolton why he had uncharacteristically published Marc's data.  "To
teach him a lesson," he responded.
 
\section{Summary}

I was privileged to have the opportunity to exploit the Parkes 210 foot radio telescope for studies ranging from solar system objects to distant radio galaxies and quasars.  I was fortunate to have been tutored by the dish-masters, John Bolton and John Shimmins.  It is to their credit that the Parkes dish has enjoyed 50 years of astronomical discovery and adventure including the exciting participation in the first manned lunar landing.

\vskip5mm

The National Radio Astronomy Observatory is a facility of the National Science Foundation operated under cooperative agreement by Associated Universities, Inc.

\vskip5mm

{\it Facilities:} \facility{Parkes}.

\end{document}